\documentclass[notitlepage,prd,superscriptaddress,onecolumn,nofootinbib,showpacs]{revtex4-1}
\usepackage{amsfonts,amssymb,amsmath}
\usepackage{mathrsfs}
\usepackage{color}
\usepackage{graphicx}
\usepackage{dcolumn}
\usepackage{bm}
\usepackage{pslatex}
\usepackage[T1]{fontenc}
\usepackage[latin1]{inputenc}
\usepackage{graphicx}
\usepackage{epsfig}
\usepackage{longtable}
\usepackage{float}
\usepackage{calc}
\usepackage{ifthen}

\begin{document}
\title{Clockwork Higgs portal model for freeze-in dark matter}
\author{Jinsu Kim}
\email{kimjinsu@kias.re.kr}
\affiliation{Quantum Universe Center, Korea Institute for Advanced Study, Seoul 02455, Korea}
\author{John McDonald}
\email{j.mcdonald@lancaster.ac.uk}
\affiliation{Department of Physics, Lancaster University, Lancaster LA1 4YB, United Kingdom}

\begin{abstract}

The clockwork mechanism can explain interactions which are dimensionally very weak without the need for very large mass scales. We present a model in which the clockwork mechanism generates the very small Higgs portal coupling and dark matter particle mass necessary to explain cold dark matter via the freeze-in mechanism. We introduce a TeV-scale scalar clockwork sector which couples to the Standard Model via the Higgs portal. The dark matter particle is the lightest scalar of the clockwork sector. We show that the freeze-in mechanism is dominated by decay of the heavy clockwork scalars to light dark matter scalars and Higgs bosons. In the model considered, we find that freeze-in dark matter is consistent with the clockwork mechanism for global charge $q$ in the range $2 \lesssim q \lesssim 4$ when the number of massive scalars is in the range $10 \leq N \leq 20$. The dark matter scalar mass and portal coupling are independent of $q$ and $N$. For a typical TeV-scale clockwork sector, the dark matter scalar mass is predicted to be of the order of a MeV. 

\end{abstract}
\keywords{Dark matter, Clockwork, FIMP, Higgs portal}

\maketitle

\section{Introduction}
\label{sec:intro}
The clockwork mechanism \cite{Choi:2015fiu,cl1,giudice} is a way to explain the existence of interactions which are much weaker than those dimensionally expected in a theory with a characteristic mass scale. Such interactions are usually created by integrating out particles which have masses much larger than the mass scale of the low-energy effective theory. Various aspects of particle theory and cosmology are conventionally explained via interactions characterized by a large mass scale, such as neutrino masses, the axion solution to the strong {\it CP} problem, and suppressed baryon number violation. In clockwork models, the generation of very weak interactions without the need for very large mass particles may allow particle physics and cosmology to be explained entirely in terms of a TeV-scale theory.
It may also allow the naturalness of the weak scale to be understood by eliminating large quantum corrections due to heavy particles \cite{noscale}.
There have been efforts to generalize the clockwork mechanism, for example to go beyond nearest-neighbor interactions in \cite{CWgeneral} and to formulate a gauged $U(1)$ clockwork in \cite{CWgauged}. The clockwork mechanism has also been used in a number of specific applications, including neutrino masses through the seesaw mechanism \cite{CWDM, CWneutrino}, muon $g-2$ \cite{CWmuong2}, axions \cite{CWaxions}, dark matter \cite{CWDM}, composite Higgs \cite{CWComposite}, the weak gravity conjecture \cite{wgc}, and inflation \cite{CWinflation}. A critical discussion of the clockwork mechanism is given in \cite{craig}; see also \cite{reply}.

In the case of a scalar clockwork model \cite{cl1,giudice}, very weak interactions can be achieved by introducing a sector consisting of a chain of $N + 1$ fundamental fields $\pi_{j}$ which transform as the Goldstone bosons of a spontaneously broken global $U(1)^{N+1}$ symmetry. 
This symmetry is also broken explicitly to a residual spontaneously broken $U(1)$ symmetry, which leaves a single massless eigenstate, $a_{0}$. The $a_{0}$ field has a very small mixing angle in the expansion of the field at the end of the chain, $\pi_{N}$, in terms of mass eigenstates. A fundamental assumption of the clockwork model is that only the $\pi_{N}$ field couples to the Standard Model (SM) sector. In this case, the $a_{0}$ field will have highly suppressed couplings to the SM fields and will also obtain a mass much smaller than the mass scale of the clockwork sector. It is also possible to obtain the clockwork sector from discrete extra dimensions\footnote{
The assumption that only $\pi_{N}$ couples to the SM sector is conceptually similar to the assumption that SM fields exist at a particular point in extra-dimensions in brane models.
} \cite{cl1,giudice}. The clockwork sector can therefore be viewed as the implementation of a phenomenological mechanism which explains the existence of very small couplings via the sector's structure, where by structure we mean the mass terms and couplings of the scalars, which have either a naturally large value or are equal to zero.

A specific example of a model which requires a very small mass and coupling is the freeze-in model of cold dark matter \cite{jmfr,hall}.
In this model, a feebly interacting massive particle (FIMP) is produced by the decay of particles which are in thermal equilibrium\footnote{
A review of the freeze-in mechanism and FIMP models is given in \cite{tenkanen}.
}. For example, dark matter can be produced by the decay of thermal bath Higgs bosons interacting with dark matter scalars via the Higgs portal\footnote{
For a recent Higgs-portal vector dark matter model via freeze-in mechanism, see \cite{Ko:2016fcd}.
} \cite{jmfr}. The freeze-in mechanism requires that the dark matter particles are out of equilibrium, which in turn requires that the Higgs portal coupling is very small. The mass of the dark matter particle is also typically much smaller than a GeV. The clockwork mechanism is particularly well motivated as an explanation for very small couplings to $H^{\dagger}H$, as these cannot be explained by a conventional symmetry. In this paper we will present a scalar clockwork version of the Higgs portal freeze-in model which can account for the small portal coupling and dark matter particle mass.

The paper is organized as follows. In Sec. \ref{sec:CWHportal} we introduce the scalar clockwork Higgs portal model, in Sec. \ref{sec:FIMPDM} we calculate the dark matter density from freeze-in, in Sec. \ref{sec:results} we present our results, and in Sec. \ref{sec:conclusion} we discuss our conclusions.

\section{A scalar clockwork Higgs portal model}
\label{sec:CWHportal}

\subsection{The scalar clockwork sector}

The scalar clockwork sector is a sector of real scalar fields with a particular pattern of mass mixing. It can be derived as the effective theory of a spontaneously and explicitly broken global symmetry or as the low-energy limit of a theory of discrete extra dimensions \cite{cl1,giudice}. 
We will follow the approach based on a broken global symmetry.

The scalar clockwork sector can be constructed by considering a set of $N + 1$ scalars, $\pi_{j}$ ($j = 0, \cdots, N$), which are the Goldstone bosons of a $G = U(1)^{N+1} = U(1)_{0} \times U(1)_{1} \times \cdots \times U(1)_{N}$ global symmetry acting on complex fields $\phi_{j}$. The symmetry is spontaneously broken at a scale $f$, such that $\phi_{j} = fU_{j}$ where $U_{j} = \exp(i\pi_{j}/f)$. The symmetry is also explicitly broken. In the clockwork model of \cite{cl1}, the explicit symmetry-breaking term is a dimensionless product of $\phi_{j}$ fields parametrized by a coupling $\epsilon$, while in \cite{giudice} the symmetry breaking is considered to be due to spurion mass squared parameters. We will use the latter method in our construction. In this case there is a natural symmetry of the interaction terms, $\pi_{j} \leftrightarrow -\pi_{j}$, which keeps the lightest clockwork scalar stable. The charges of the spurion mass squared terms $m_{j}^{2}$ ($j=0, \cdots, N-1$) under the $U(1)_{i}$ factors of $G$ are \cite{giudice}
\begin{align}\label{eqn:1}
Q_{i}[m_{j}^{2}] = \delta_{i,j} - q\delta_{i,j+1}\,.
\end{align}
The resulting Lagrangian of the $\pi_{j}$ is then
\begin{align}\label{eqn:2}
\mathcal{L} = 
\frac{f^{2}}{2}\sum_{j=0}^{N}
\partial_{\mu}U_{j}^{\dagger}
\partial^{\mu}U_{j}
+\frac{m^{2}f^{2}}{2}\sum_{j=0}^{N-1}
\left(
	U_{j}^{\dagger}U_{j+1}^{q}
	+\text{H.c.}
\right)\,.
\end{align}
For simplicity, the values of $m_{j}^{2}$ are assumed to all equal a common symmetry-breaking spurion mass squared term, $m_{j}^{2} = m^2$. 
On expanding in $\pi_{j}/f$, Eq.~\eqref{eqn:2} becomes
\begin{align}\label{eqn:3}
	\mathcal{L} =
	\frac{1}{2}\partial_{\mu}\pi_{j}\partial^{\mu}\pi_{j}
	-V(\pi)\,,
\end{align}
where
\begin{align}\label{eqn:4}
	V(\pi) = 
	\frac{m^{2}}{2}\sum_{j=0}^{N-1}\left(
		\pi_{j} - q\pi_{j+1}
	\right)^{2}
	-\frac{m^{2}}{24f^{2}}\sum_{j=0}^{N-1}\left(
	\pi_{j} - q\pi_{j+1}
	\right)^{4}
	+\mathcal{O}(\pi^{6})\,.
\end{align}
This potential explicitly breaks $G$ to a single residual spontaneously broken global $U(1)$. On diagonalizing the resulting mass matrix, the mass eigenstate scalars $a_{j}$  are related to the  $\pi_{j}$ via \cite{giudice}
\begin{align}\label{eqn:5}
	\pi_{j} = O_{ji}a_{i}\,,
\end{align}
where
\begin{align}\label{eqn:6}
	O_{j0} = \frac{\tilde{N}_{0}}{q^{j}}
	\,,\qquad
	O_{jk} = 
	\tilde{N}_{k} \left[ q \sin \left(\frac{jk\pi}{N+1}\right) - \sin \left( \frac{ \left(j + 1 \right) k \pi}{N+1} \right) \right]\,.
\end{align}
Here $i,\,j = 0, \cdots , N$ and $k = 1, \cdots , N$. $\tilde{N}_{0}$  and $\tilde{N}_{k}$ are given by
\begin{align}\label{eqn:7}
	\tilde{N}_{0} = 
	\sqrt{ \frac{q^2 - 1}{q^2 - q^{-2N}} }
	\,,\qquad
	\tilde{N}_{k} = \sqrt{\frac{2}{\left(N+1\right) \lambda_{k}}}
	\,,
\end{align}
where
\begin{align}\label{eqn:8}
	\lambda_{k} = q^2 + 1 - 2 q \cos \left(\frac{k \pi}{N+1} \right)\,.
\end{align}
The masses of the mass eigenstate scalars are
\begin{align}\label{eqn:9}
	m_{a_{0}}^2 = 0\,,\qquad
	m_{a_{k}}^2 = \lambda_{k} m^2
	\,.
\end{align}
In particular, for large $N$ values we have $m_{a_{1}} = (q-1) m$ and $m_{a_{N}} = (q + 1) m$.

The important feature is the massless scalar $a_{0}$, which is the Goldstone boson associated with the residual spontaneously broken $U(1)$.
Since $a_{0}$ is a Goldstone boson, it does not appear in the potential \eqref{eqn:4} due to the shift symmetry of $a_{0}$. 
The clockwork mechanism is based on the fundamental assumption that only the $\pi_{N}$ field interacts with SM fields.
In this case the $a_{0}$ scalar will have highly suppressed couplings to the SM sector due to the $q^{-N}$ factor in $O_{N0}$ if $q > 1$ and $N$ is sufficiently large compared to 1. 

In the clockwork Higgs portal model, the coupling of $\pi_{N}$ to the SM is assumed to be via the Higgs portal. For example, this can be achieved\footnote{
Here we are choosing to construct the Higgs portal interaction by using an $|H|^{2}$ dependent spurion mass term and the $U_{N}$ factor. Alternatively, we could simply construct the interaction directly by coupling $\pi_{N}^{2}$ to $|H|^{2}$, similar to the construction of the axion portal interaction given by Eq.~(2.21) of \cite{giudice}.
} by introducing a further spurion mass term $m_{N}^2(1 + |H|^{2}/\Lambda^2)$, which transforms as $Q_{N}[m_{N}^2(1 + |H|^{2}/\Lambda^2)] = q_N$, $Q_{j}[m_{N}^2(1 + |H|^{2}/\Lambda^2)] = 0$, $j = 0, \cdots ,\,N-1$. We then introduce an additional term given by
\begin{align}\label{eqn:10}
	\mathcal{L} \supset
	\frac{m_{N}^{2} f^{2}}{2}\left(
		1 + \frac{|H|^{2}}{\Lambda^{2}}
	\right)\left(
		U_{N}^{q_N \; \dagger} + \text{H.c.}
	\right)\,,
\end{align}
where $H$ is the SM Higgs doublet.
(For simplicity we will set $q_N= 1$.) In addition to coupling $\pi_{N}$ to the SM Higgs boson, this term also explicitly breaks the residual $U(1)$ symmetry, which allows $a_{0}$ to couple to the other clockwork scalars in the potential. Since the only mass scale in the theory prior to explicit symmetry breaking is $f$, we will consider $\Lambda \approx f$ in the following, although in general $\Lambda$ could be different from $f$. 

Expanding $U_{N}$ in terms of $\pi_{N}$ then gives
\begin{align}\label{eqn:11}
	\mathcal{L} &\supset
	\frac{m_{N}^2 f^2}{2}\left(1 + \frac{|H|^2}{\Lambda^{2}} \right) \left(2 - \frac{\pi_{N}^{2}}{f^{2}} + \frac{1}{12} \frac{\pi_{N}^{4}}{f^{4}} + \cdots \right)
	\nonumber\\
	&=
	m_{N}^{2}f^{2}
	+\frac{m_{N}^{2}f^{2}}{\Lambda^{2}}|H|^{2}
	-\frac{m_{N}^{2}}{2}\pi_{N}^{2}
	-\frac{m_{N}^{2}}{2\Lambda^{2}}|H|^{2}\pi_{N}^{2}
	+\frac{m_{N}^{2}}{24f^{2}}\pi_{N}^{4}
	+\frac{m_{N}^{2}}{24f^{2}\Lambda^{2}}|H|^{2}\pi_{N}^{4}
	+\cdots\,.
\end{align}
We will work in the unitary gauge and write $|H|^2 = (h+v)^{2}/2$, with $v = 246$ GeV being the vacuum expectation value of the SM Higgs.
Then from Eq.~\eqref{eqn:11} we obtain
\begin{align}\label{eqn:12}
	\mathcal{L} \supset
	- \frac{m_{N}^2}{2} \left(1 + \frac{v^2}{2 \Lambda^2} \right) \pi_{N}^2 - \frac{m_{N}^{2}v}{2 \Lambda^2} h \pi_{N}^{2} - \frac{m_{N}^{2}}{4 \Lambda^2} h^2 \pi_{N}^{2} + \frac{m_{N}^{2}}{24f^{2}}\left(1 + \frac{v^2}{2 \Lambda^2} \right)\pi_{N}^{4} + \cdots\,,
\end{align}
where the center dots contain terms coming from higher-order interactions such as $|H|^{2}\pi_{N}^{4}$.

The symmetry-breaking approach results in a scalar clockwork sector which includes higher-order nonrenormalizable interactions in the potential. There will also be derivative interactions between the $\pi_{j}$ fields, of the form $(\partial^{\mu} \pi_{j} \partial_{\mu} \pi_{j})^{2}/f^{4}$,  from integrating out the radial fields $\eta_{j}$ of the complex scalars $\phi_{j} \equiv (\eta_{j} + f)e^{i \pi_{j}/f}/\sqrt{2}$.   
However, in order to explain small masses and couplings, the clockwork mechanism requires only the lowest-order terms of the effective theory. Therefore, we can also consider a minimal clockwork model based on a renormalizable sector which has only canonical kinetic terms and a renormalizable potential. In the renormalizable limit, the model becomes
\begin{align}\label{eqn:12a}
    {\cal L} = \frac{1}{2} \partial_{\mu} \pi_{j} \partial^{\mu} \pi_{j} - V^{{\rm ren}}
  \,,
\end{align}
where
\begin{align}\label{eqn:12b}
V^{{\rm ren}} = \frac{m^{2}}{2}\sum_{j=0}^{N-1}\left(
		\pi_{j} - q\pi_{j+1}
	\right)^{2}
	+ g_{1} \sum_{j=0}^{N-1}\left(
	\pi_{j} - q\pi_{j+1}
	\right)^{4}
+ \frac{m_{\pi_{N}}^2}{2} \pi_{N}^2 + g_{2} h \pi_{N}^{2} + g_{3} h^2 \pi_{N}^{2}  + g_{4} \pi_{N}^{4}
  \,.
\end{align}
In the case of the symmetry-breaking model, the renormalizable mass and coupling terms are given by
\begin{align}\label{eqn:12c}
g_{1} = -\frac{m^{2}}{24f^{2}}\;,\;\;\; g_{2} = \frac{m_{N}^{2}v}{2 \Lambda^2}\;,\;\;\; g_{3} =  \frac{m_{N}^{2}}{4 \Lambda^2}\;,\;\;\;  g_{4} = - \frac{m_{N}^{2}}{24f^{2}}\left(1 + \frac{v^2}{2 \Lambda^2} \right) \;, \;\;\; m_{\pi_{N}}^{2} = m_{N}^2 \left(1 + \frac{v^2}{2 \Lambda^2} \right)
  \,.
\end{align}
In general, a renormalizable sector could arise from a fundamental theory in the same way as the renormalizable SM itself, with an UV completion at a common scale (such as the Planck scale). In this case the $U(1)$ symmetries are replaced by the corresponding shift symmetries of the $\pi_{j}$ fields. The first two terms in Eq.~\eqref{eqn:12b} explicitly break the $N$ shift symmetries to a single residual shift symmetry. The remaining terms introduce interactions between $\pi_{N}$ and the SM Higgs and break the residual shift symmetry. In the following we will consider both the full symmetry-breaking model clockwork sector and the renormalizable limit of the clockwork sector.

\subsection{Mass eigenstates and Higgs portal interactions}

The $\pi_{N}^{2}$ [$\equiv (\sum_{j = 0}^{N} O_{Nj} a_{j})^{2}$] term in Eq.~\eqref{eqn:12a} will cause a mass mixing between $a_{0}$ and $a_{k}$ ($k = 1, \cdots, N$) which is proportional to $O_{N0} O_{Nk}$. (In the following we will assume that $v^{2}/2 \Lambda^2 \ll 1$ and so set the $\pi_{N}^2$ term to $-m_{N}^{2} \pi_{N}^{2}/2$ for simplicity.) In general, it is difficult to diagonalize the mass matrix to obtain the mass eigenvalues and eigenvectors. To obtain a useful expression which will allow us to calculate the freeze-in dark matter density, we adopt the following approach. Once the $m_{N}^{2}$ term is introduced, the mass mixing term between $a_{0}$ and $a_{k}$ is given by $2 m_{N}^{2} O_{N0} a_{0} \sum_{k = 1}^{N} O_{Nk} a_{k}$. Therefore, only the linear combination proportional to  $\sum_{k = 1}^{N} O_{Nk} a_{k}$ will couple to $a_{0}$. Prior to introducing $m_{N}^{2}$ term, we will assume that the mass terms are close to degenerate for $k = 1, \cdots,\,N$, i.e. $m_{a_{1}}^2 \approx m_{a_{2}}^{2} \approx \cdots \approx m_{a_{N}}^{2}$.  (We will refer to this as the degenerate mass approximation.) In this case we will set all the diagonal terms to $m_{a_{1}}^2$. In practice the masses $m_{a_{k}}$ will be spread over a relatively small range $\Delta m$, where $m_{a_{N}}\approx m_{a_{1}}+\Delta m$ and $\Delta m = 2m_{a_{1}}/(q-1) \lesssim m_{a_{1}}$. Therefore, we expect the degenerate mass approximation to provide a good estimate of the contribution of the heavy mass eigenstate scalars to the freeze-in dark matter density.

In the degenerate mass approximation, the mass matrix for the $a_{k}$ scalars prior to introducing the $\pi_{N}^2$ mass term is simply $m_{a_{1}}^{2}$ times the identity matrix. Therefore, we can make an arbitrary orthogonal transformation of the $a_{k}$ fields to a new mass eigenstate basis $a_{k}^{*}$. Thus we can choose a new basis such that $a_{1}^{*} = K\sum_{k=1}^{N} O_{Nk} a_{k}$, where $K$ is a normalization factor which satisfies $K^{2} \sum_{k=1}^{N} O_{Nk}^{2} = 1$. Since $O_{N0}^{2} + \sum_{k=1}^{N} O_{Nk}^{2} = 1$, the normalization factor $K$ is given by $K^{2} = 1/(1 - O_{N0}^{2})$. Since $O_{N0}^{2} \ll 1$ [see Eq.~\eqref{eqn:6}], it follows that $K \approx 1$. In this basis the $\pi_{N}$ field is given by $\pi_{N} = O_{N0} a_{0} + \sum_{k = 1}^{N} O_{Nk} a_{k} \approx O_{N0} a_{0} + a_{1}^{*}$. 

Thus the Higgs portal interaction of the heavy clockwork scalars in the degenerate mass approximation reduces to a system of two scalars, $a_{0}$ and $a_{1}^{*}$, with $a_{2}^{*}$ to $a_{N}^{*}$ decoupled from the Higgs portal. Once the $\pi_{N}^2$ mass term is introduced, the mass terms of the ($a_{0}$, $a_{1}^{*}$) system become
\begin{align}\label{eqn:14}
	-\frac{1}{2} \overline{m}_{a_{0}}^{2} a_{0}^{2} - \overline{m}_{a_{0} a_{1}^{*}}^{2} a_{0} a_{1}^{*} - \frac{1}{2} \overline{m}_{a_{1}^{*}}^{2} a_{1}^{*\,2}
	\,,
\end{align}
where we have defined $\overline{m}_{a_{0}}^{2} = m_{N}^{2} O_{N 0}^{2}$,  $\overline{m}_{a_{0} a_{1}^{*}}^{2} = m_{N}^{2} O_{N 0}$ and $\overline{m}_{a_{1}^{*}}^{2} = m_{N}^{2} + m_{a_{1}}^{2}$. 
Diagonalizing the mass matrix results in mass eigenstates $\hat{a}_{0}$ and $\hat{a}_{1}$, which are related to $a_{0}$ and $a_{1}^{*}$ by
\begin{align}\label{eqn:15}
	a_{0} = \hat{a}_{0}\cos\alpha
	+\hat{a}_{1}\sin\alpha
	\,,\qquad
	a_{1}^{*} = -\hat{a}_{0}\sin\alpha
	+\hat{a}_{1}\cos\alpha
	\,,
\end{align}	
where the mixing angle $\alpha$ is given by
\begin{align}\label{eqn:16}
	\tan(2 \alpha) = \frac{2 \overline{m}_{a_{0} a_{1}^{*}}^{2} }{ \overline{m}_{a_{1}^{*}}^{2} 
	- \overline{m}_{a_{0}}^{2}  }
	\,.
\end{align}
Since $O_{N0} \ll 1$, we can assume that $\overline{m}_{a_{0}}^{2} \ll \overline{m}_{a_{0} a_{1}^{*}}^{2} \ll \overline{m}_{a_{1}^{*}}^{2} $. In this limit $\alpha$ is given by
\begin{align}\label{eqn:17}
	\alpha \approx \frac{\overline{m}_{a_{0}a_{1}^{*}}^{2}}{\overline{m}_{a_{1}^{*}}^{2}} = \frac{O_{N0} m_{N}^{2}}{m_{N}^{2}  + m_{a_{1}}^{2}}
	\,,
\end{align}
where $\alpha \ll 1$ since $O_{N0} \ll 1$. The mass eigenstates are then
\begin{align}\label{eqn:18}
	\hat{a}_{0} \approx a_{0} - \alpha \, a_{1}^{*}
	\,,\qquad
	\hat{a}_{1} \approx a_{1}^{*}+\alpha \, a_{0}
	\,.
\end{align}
The corresponding mass eigenvalues are, using $\alpha \ll 1$,
\begin{align}\label{eqn:19}
	m_{\hat{a}_{0}} \approx
	\gamma_{0}^{1/2} O_{N0}m_{N}
	\,,\qquad
	m_{\hat{a}_{1}} \approx
	\overline{m}_{a_{1}^{*}}
	= \sqrt{m_{N}^{2} + m_{a_{1}}^{2}}
	\,,
\end{align}
where $\gamma_{0} =  m_{a_{1}}^{2}/\left(m_{N}^{2} + m_{a_{1}}^{2} \right)$. In terms of the mass eigenstates, the $\pi_{N}$ expansion is
\begin{align}\label{eqn:20}
	\pi_{N} \approx O_{N0} a_{0} + a_{1}^{*} \approx \gamma_{0} O_{N0} \hat{a}_{0}
	+ \hat{a}_{1}
	\,.
\end{align}
The leading order interaction terms between $\hat{a}_{0}$, $\hat{a}_{1}$, and the Higgs boson $h$ are then given by
\begin{align}\label{eqn:21}
	V^{{\rm int}} = \frac{m_{N}^{2}}{2 \Lambda^{2}} \left(vh + \frac{h^{2}}{2} \right) \pi_{N}^{2}   \equiv \lambda_{1} h \hat{a}_{0}^{2} + \lambda_{2} h^{2} \hat{a}_{0}^{2} + \lambda_{3} h \hat{a}_{0} \hat{a}_{1} + \lambda_{4} h^{2} \hat{a}_{0} \hat{a}_{1} + \lambda_{5} h \hat{a}_{1}^{2} + \lambda_{6} h^{2} \hat{a}_{1}^{2} 
	\,,
\end{align}
where
\begin{align}\label{eqn:22}
\lambda_{1} = \frac{m_{N}^{2}}{2 \Lambda^{2}} v \gamma_{0}^{2} O_{N0}^{2}
\,,\quad
\lambda_{2} = \frac{m_{N}^{2}}{4 \Lambda^{2}}\gamma_{0}^{2} O_{N0}^{2}
\,,\quad
\lambda_{3} = \frac{m_{N}^{2}}{\Lambda^{2}} v \gamma_{0}O_{N0}
\,,\quad
\lambda_{4} = \frac{m_{N}^{2}}{2 \Lambda^{2}} \gamma_{0}O_{N0}
\,,\quad 
\lambda_{5} = \frac{m_{N}^{2}}{2 \Lambda^{2}} v 
\,,\quad
\lambda_{6} = \frac{m_{N}^{2}}{4 \Lambda^{2}}
\,.\quad
\end{align}
Note that the $\hat{a}_{1}$ scalars will be kept in thermal equilibrium via the interactions $\lambda_{5}$ and $\lambda_{6}$, which have no large suppression factor. Similarly, the other heavy scalars $a_{2}^{*}, \cdots, a_{N}^{*}$ will be kept in thermal equilibrium due to their interaction with $\hat{a}_{1}$ via the quartic terms in Eq.~\eqref{eqn:4}.

We note that since the shift symmetry of the Goldstone boson field $a_{0}$ is broken only by the portal interaction \eqref{eqn:10}, the couplings of $\hat{a}_{0} \, (\approx a_{0})$ will always have a factor of $O_{N0}$ for each $\hat{a}_{0}$. Therefore, quantum corrections to the portal couplings in Eq.~\eqref{eqn:21}, which are logarithmic with cutoff $f$, will be proportional to the same $O_{N0}$ factors as the tree-level couplings and so will be small compared to the tree-level portal couplings.
The portal couplings of the heavy clockwork scalars $a_{1}$, $...$, $a_{N}$ are not strongly suppressed, since the mixing angles $O_{Nj}$ in $\pi_{N}$ are not very small for $j = 1$, $...$, $N$.  The logarithmic quantum corrections to these couplings are necessarily proportional to the tree-level portal couplings, since in their absence the clockwork sector would be completely decoupled from the SM sector. Therefore, quantum corrections to the Higgs portal couplings of $a_{1}$, $...$, $a_{N}$ will also be small compared to the tree-level portal couplings.\footnote{The absence of large quantum corrections to the portal couplings of $a_{0}$, $...$, $a_{N}$ is equivalent to the effect of the Higgs portal couplings of $\pi_j$ ($j = 0$, $...$, $N$) generated by quantum corrections being small.} Thus the freeze-in mechanism is generally stable with respect to quantum corrections.

We next apply the interaction \eqref{eqn:21} to calculate the relic density of $\hat{a}_{0}$ dark matter and to determine the conditions on the clockwork model necessary to account for the observed density of dark matter.

\section{Freeze-in density of $\hat{a}_{0}$ dark matter}
\label{sec:FIMPDM}

The freeze-in mechanism for the production of out-of-equilibrium dark matter \cite{jmfr,hall} is based on the accumulation of dark matter particles produced by the decay of a particle which is in thermal equilibrium. This was first considered for the case of Higgs boson decay to dark matter scalars in \cite{jmfr} and later generalized in \cite{hall}. For the case of a scalar particle $B_{1}$ decaying to a pair of scalars $B_{2}$ and $X$, where $X$ is the FIMP dark matter particle, the yield of $X$ particles from freeze-in is \cite{hall}
\begin{align}\label{eqn:23}
	Y_{X} = \frac{405 \sqrt{10} }{8 \pi^{4}} \frac{g_{B_{1}} \Gamma_{B_{1}} M_{P}}{m_{B_{1}}^{2} g_{*S} \sqrt{g_{*}}  }
	\,,
\end{align}
where $M_{P}$ is the reduced Planck mass, $m_{B_{1}}$ is the $B_{1}$ scalar particle mass, $\Gamma_{B_{1}}$ is the partial decay width of $B_{1} \rightarrow B_{2} X$. $g_{B_{1}}$, $g_{*S}$, and $g_{*}$ are respectively internal degrees of freedom of $B_{1}$, the effective degrees of freedom in the thermal bath for the entropy, and the effective degrees of freedom for the energy density.
We will consider $g_{*} = g_{*S}$ in the following. Most of the $X$ production occurs at $T \sim m_{B_{1}}$, so we consider $g_{*}$ to be equal to its value at $T \approx m_{B_{1}}$. In practice $g_{*} = 106.75$, corresponding to the fields of the SM.
We will also consider the decaying particle to be a real scalar, so that $g_{B_{1}} = 1$. Then the present $X$ dark matter density is
\begin{align}\label{eqn:24}
	\Omega_{X,0}h^{2} \approx
	1.1 \times 10^{27} \frac{1}{g_{*}^{3/2}} 
	\frac{m_{X} \Gamma_{B_{1}}}{m_{B_{1}}^{2}}
	\,.
\end{align}

In the analysis in this paper we will consider $m_{\hat{a}_{1}} > m_{h}$. In this case freeze-in via the Higgs portal interaction is due to the process $\hat{a}_{1} \rightarrow h \hat{a}_{0}$. The $\hat{a}_{1}$ decay rate is given by
\begin{align}\label{eqn:25}
	\Gamma_{\hat{a}_{1}} = \frac{\lambda_{3}^{2}}{16 \pi m_{\hat{a}_{1}}}
	\,.
\end{align}
Therefore, from Eq.~\eqref{eqn:24} and with $X \equiv \hat{a}_{0}$ and $B_{1} \equiv \hat{a}_{1}$, we find
\begin{align}\label{eqn:26}
	\Omega_{\hat{a}_{0}} h^{2} \approx 1.1 \times 10^{27} \frac{1}{g_{*}^{3/2}} \frac{ m_{\hat{a}_{0}} \lambda_{3}^{2}}{16 \pi m_{\hat{a}_{1}}^{3}}
	\,.
\end{align}
Thus the condition for $\hat{a}_{0}$ from freeze-in to be able to account for the observed dark matter density is
\begin{align}\label{eqn:27}
	\lambda_{3} \approx 2.1 \times 10^{-13} \left( \Omega_{\hat{a}_{0}} h^{2} \right)^{1/2} g_{*}^{3/4} \frac{m_{\hat{a}_{1}}^{3/2}}{m_{\hat{a}_{0}}^{1/2}}
	\,.
\end{align}
Replacing $\lambda_{3}$, $m_{\hat{a}_{1}} [= m_{N} \left(1 - \gamma_{0}\right)^{-1/2} ]$, and $m_{\hat{a}_{0}}$ by their expressions in terms of model parameters and mixing angles [Eqs.~\eqref{eqn:22} and \eqref{eqn:19}], this condition becomes
\begin{align}\label{eqn:28}
	\frac{m_{N}v}{\Lambda^{2}}
	\gamma_{0}^{5/4}(1-\gamma_{0})^{3/4}O_{N0}^{3/2}
	\approx
	2.4\times 10^{-12}\left(
		\frac{\Omega_{\hat{a}_{0}}h^{2}}{0.12}
	\right)^{1/2}\left(
		\frac{g_{*}}{106.75}
	\right)^{3/4}
	\,.
\end{align}
Using the definition of $\gamma_{0} \equiv m_{a_{1}}^{2}/(m_{N}^{2}+m_{a_{1}}^{2})$, we find
\begin{align}\label{eqn:29}
	\gamma_{0}^{5/4}(1-\gamma_{0})^{3/4} = \frac{m_{a_{1}}^{5/2}m_{N}^{3/2}}{(m_{N}^{2}+m_{a_{1}}^{2})^{2}}\,.
\end{align}
We also assume that $q$ is large compared to 1 and $N$ is significantly larger than 1, which will be true for realistic clockwork sectors.
In this case $\tilde{N}_{0} \approx 1$ [see Eq.~\eqref{eqn:7}]. From Eqs.~\eqref{eqn:28}, \eqref{eqn:29}, and \eqref{eqn:6}, we then obtain 
\begin{align}\label{eqn:30}
	q^{-3N/2}
	\approx
	9.8 \times 10^{-15}\left(
		\frac{\Omega_{\hat{a}_{0}}h^{2}}{0.12}
	\right)^{1/2}\left(
		\frac{g_{*}}{106.75}
	\right)^{3/4}\left(
		\frac{246\,{\rm GeV}}{v}
	\right)
	\frac{\Lambda^{2}(m_{N}^{2}+m_{a_{1}}^{2})^{2}}{m_{N}^{5/2}m_{a_{1}}^{5/2}}\,{\rm GeV}^{-1}
	\,.
\end{align}
Thus the condition for freeze-in via the clockwork Higgs portal to account for the observed density of dark matter is
\begin{align}\label{eqn:32}
	\ln q \approx
	\frac{2}{3N}\left[
		20.7
		-\ln \beta
		-2\ln\left(
			\frac{\Lambda}{10\,{\rm TeV}}
		\right)
		+\ln\left(
			\frac{m_{a_{1}}}{1\,{\rm TeV}}
		\right)
		-\ln\left[
			\frac{(1+m_{N}^{2}/m_{a_{1}}^{2})^{2}}{(m_{N}/m_{a_{1}})^{5/2}}
		\right]
	\right]\,,
\end{align}
where 
\begin{align}\label{eqn:33}
	\beta \equiv
	\left(
	\frac{\Omega_{\hat{a}_{0}}h^{2}}{0.12}
	\right)^{1/2}\left(
	\frac{g_{*}}{106.75}
	\right)^{3/4}\left(
	\frac{246\,{\rm GeV}}{v}
	\right)\, .
\end{align}
(We will set $\beta = 1$ in the following.) 

The $\hat{a}_{1}$ scalars will freeze out of chemical equilibrium once their annihilation to Higgs bosons freezes out. We should therefore check that the relic $\hat{a}_{1}$ scalars can decay to $\hat{a}_{0} + h$ before nucleosynthesis. 
This requires that $\Gamma_{\hat{a}_{1}} \gtrsim H(T)$ at $T_{{\rm nuc}}$. This is satisfied if
\begin{align}\label{eqn:33a}
\lambda_{3} \gtrsim \left( \frac{16 \pi m_{\hat{a}_{1}} }{M_{P}} \right)^{1/2} \left(\frac{T_{{\rm nuc}}}{v} \right) \, v \approx  
6.1 \times 10^{-12} \, \left( \frac{m_{\hat{a}_{1}} }{1\, {\rm TeV}} \right)^{1/2} 
\left( \frac{T_{{\rm nuc}}}{10\, {\rm MeV}} \right) \, v  
\,.
\end{align}
We will see that this is easily satisfied by the values of $\lambda_{3}$ necessary to account for the observed dark matter density.

In the limit where the dark matter scalars are exactly degenerate in mass prior to mixing, the scalars $\hat{a}_{2}, \cdots, \hat{a}_{N}$ are decoupled from the Higgs portal and so would not be able to decay to $h$ plus $\hat{a}_{0}$. However, this is simply an artifact of the degenerate mass approximation, which is simply a way to estimate the total freeze-in production of dark matter scalars from the decay of the heavy clockwork scalars. We can allow a small breaking of the degeneracy which is sufficient to allow the heavy scalars to decay harmlessly via the Higgs portal without significantly altering the results of the degenerate mass approximation. In this case the decay of $\hat{a}_{2}, \cdots, \hat{a}_{N}$ to $h$ plus $\hat{a}_{0}$ also contributes to the total freeze-in density, but with a much smaller contribution than that from $\hat{a}_{1}$ decay.

\section{Results}
\label{sec:results}

\begin{figure}[h]
	\centering
	\includegraphics[width=0.45\textwidth]{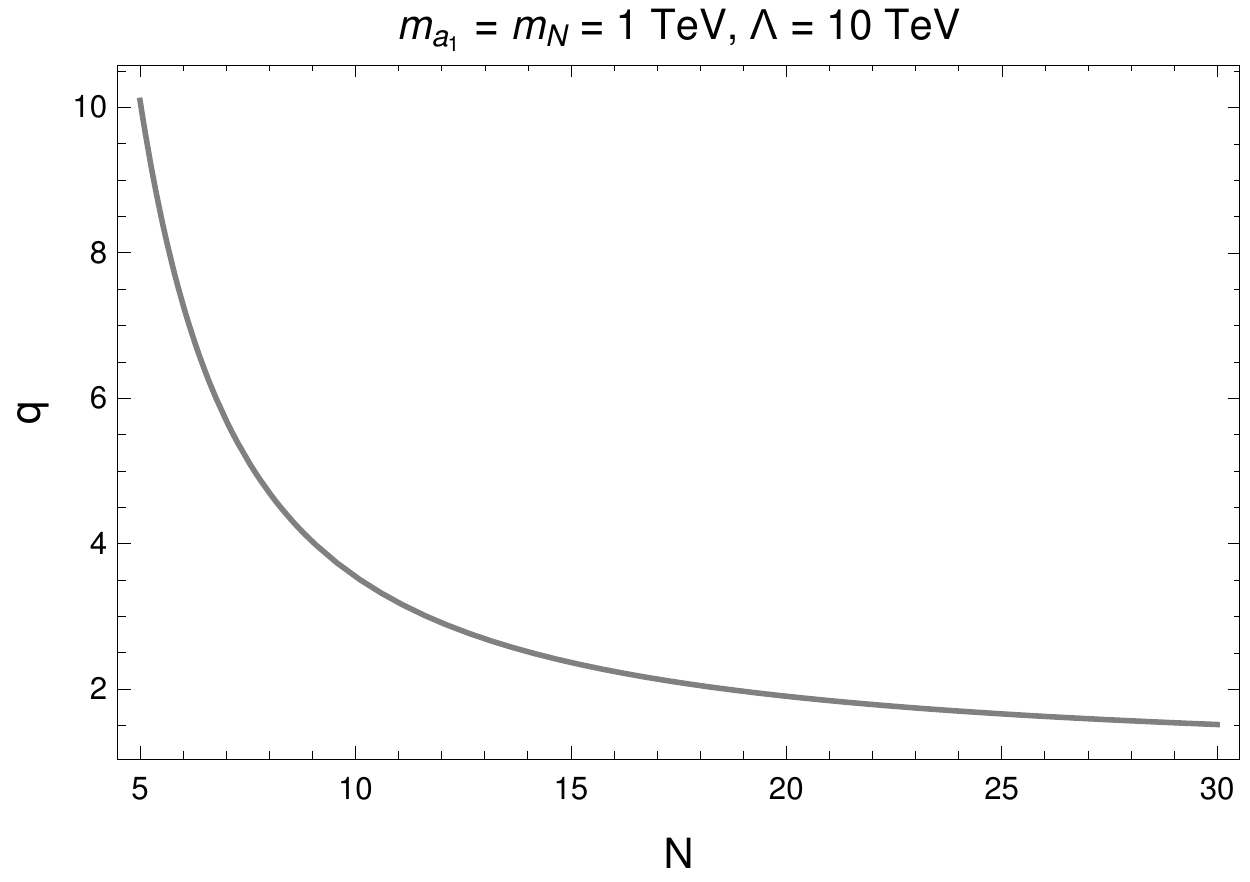}
	\caption{\label{fig:qvsN}
		Values of $q$ versus $N$ for the case $m_{a_{1}} = m_{N} = 1 \, {\rm TeV}$ and $\Lambda = 10 \, {\rm TeV}$. The solid line represents the values of $q$ and $N$ satisfying Eq.~\eqref{eqn:32}.
	}
\end{figure}

In Fig. 1 we show values of $q$ versus $N$ for the case $m_{a_{1}} = m_{N} = 1 $ TeV and $\Lambda = 10$ TeV. In this case $m_{\hat{a}_{1}} = 1.4$ TeV. We find that $q = 3.62$ for a clockwork sector with 10 massive scalars, and $q=1.90$ for a clockwork sector with 20 massive scalars.\footnote{Values of $q$ may most naturally be an integer or fractional. For a given integer $N$, $q$ can be adjusted to an integer or simple fraction by varying $m_{N}/\Lambda$ appropriately.} [The corresponding values of $m \, (= m_{a_{1}}/(q-1))$ are 0.38 and 1.1 TeV, respectively.] These values of $q$ appear to be quite reasonable and show that the clockwork mechanism can naturally generate the necessary light dark matter particle mass and very small coupling required by the freeze-in mechanism. 

An interesting feature of clockwork Higgs portal freeze-in is that a unique value of $m_{\hat{a}_{0}}$ and $\lambda_3$ is predicted for a given $\Lambda$, $m_{a_{1}}$, and $m_N$, which is independent of $q$ and $N$. This is because in this case  $m_{\hat{a}_{0}}$ and $\lambda_{3}$ are both determined by the value of $O_{N0}$. Once $O_{N0}$ is fixed by the relic dark matter density, the values of $m_{\hat{a}_{0}}$ and $\lambda_{3}$ are also fixed, independently of $q$ and $N$. This is quite different from a general freeze-in model, where larger values of the dark matter particle mass can be accommodated by simply reducing the Higgs portal coupling and so the number density of produced dark matter particles. Using Eqs.~\eqref{eqn:28}, \eqref{eqn:19}, and \eqref{eqn:22}, the values of $O_{N0}$, $m_{\hat{a}_{0}}$, and $\lambda_3$ necessary to account for dark matter are
\begin{align}\label{eqn:34}
	O_{N0} \approx
	\frac{9.8 \times 10^{-7}}{\gamma_{0}^{4/3}}
	\left(
		\frac{\Lambda}{10\,{\rm TeV}}
	\right)^{4/3}\left(
		\frac{1\,{\rm TeV}}{m_{N}}
	\right)^{5/3}\left(
		\frac{m_{a_{1}}}{1\,{\rm TeV}}
	\right)\,,
\end{align}
\begin{align}\label{eqn:35}
	m_{\hat{a}_{0}} \approx
	\frac{0.98\,{\rm MeV}}{\gamma_{0}^{5/6}}
	\left(
	\frac{\Lambda}{10\,{\rm TeV}}
	\right)^{4/3}\left(
	\frac{1\,{\rm TeV}}{m_{N}}
	\right)^{2/3}\left(
	\frac{m_{a_{1}}}{1\,{\rm TeV}}
	\right)\,,
\end{align}
and
\begin{align}\label{eqn:36}
	\lambda_{3} \approx
	\frac{9.8\times 10^{-9}}{\gamma_{0}^{1/3}}v
	\left(
		\frac{10\,{\rm TeV}}{\Lambda}
	\right)^{2/3}\left(
		\frac{m_{N}}{1\,{\rm TeV}}
	\right)^{1/3}\left(
		\frac{m_{a_{1}}}{1\,{\rm TeV}}
	\right)\,.
\end{align}
Thus the model predicts FIMP dark matter with a mass which is typically close to 1 MeV. 

Our calculation assumes that the range of heavy scalar masses is not large, so that treating them as degenerate is a reasonable approximation for determining the $\hat{a}_{0}$ density due to the decay of the heavy clockwork scalars. (In general, the freeze-in density is the sum of the contributions of the decay of each heavy scalar mass eigenstate $\hat{a}_{k}$ and so is effectively a sum of independent freeze-in processes.) The range of mass over which the $N$ heavy scalars are spread corresponds to $\Delta m = m_{a_{N}} - m_{a_{1}} \approx 2 m_{a_{1}}/(q-1)$.
For $N = 10$, $q = 3.62$, and $m_{a_{1}} = m_{N} =1\,{\rm TeV}$, we find that $\Delta m =  0.76 \,{\rm TeV}$. Therefore, $\Delta m/ m_{a_{k}} < 1$ in this case and we expect the degenerate mass approximation to be accurate up to $\mathcal{O}(1)$ correction factors in the $\hat{a}_{0}$ mass and portal coupling.
This can be seen from Eqs.~\eqref{eqn:35} and \eqref{eqn:36}, which show that $m_{\hat{a}_{0}}$ and $\lambda_{3}$ are both linear in $m_{a_{1}}$. Therefore, if we replace $m_{a_{1}}$  by $m_{a_{1}} + \Delta m \equiv m_{a_{N}}$, we will obtain upper limits on $m_{\hat{a}_{0}}$ and $\lambda_{3}$, with the true values expected to lie between those with $m_{a_{k}} = m_{a_{1}}$ and $m_{a_{k}} = m_{a_{N}}$ for all value of $k$. Similarly, for $N = 20$ and $q=1.90$, we obtain $\Delta m =  2.21 \,{\rm TeV}$. Therefore, $\Delta m/m_{a_{k}} \sim 1$ and so we would again expect the degenerate mass approximation to give a reasonable estimate of the $\hat{a}_{0}$ mass and portal coupling. The degenerate mass approximation has the great advantage of producing analytical results. To achieve greater accuracy, we would need to perform a full numerical diagonalization of the mass matrix for each set of model parameters. 
  
In general, in models where the clockwork sector arises as the effective theory of an UV completion, nonrenormalizable interactions would also be expected. Of these, the most important for the freeze-in model are higher-order derivative interactions of the generic form 
\begin{align}\label{eqn:36aa}
 \frac{1}{\Lambda_{UV}^{4}} 
\partial_{\mu} \pi_{i} \partial^{\mu} \pi_{j} 
\partial_{\nu} \pi_{k} \partial^{\nu} \pi_{l}  
  \,,
\end{align}
where $\Lambda_{UV}$ is the scale of the UV completion (with $\Lambda_{UV} \sim f$ for the symmetry-breaking model). These result in interactions between $a_{0}$ and $a_{k}$ which have no large suppression from $O_{N0}$ factors. For example, 
\begin{align}\label{eqn:36a}
\frac{1}{\Lambda_{UV}^{4}} (\partial_{\mu} \pi_{1} \partial^{\mu} \pi_{1})^{2} \rightarrow \frac{1}{\Lambda_{UV}^{4}} O_{10} O_{1j}^{3} (\partial_{\mu} a_{0} \partial^{\mu} a_{j}) (\partial_{\nu} a_{j} \partial^{\nu} a_{j})  
\,,
\end{align} 
where the product $O_{10}O_{1j}^{3}$ gives the smallest suppression of this class of operator when $j = N/2$. Since the $\hat{a}_{0} (\approx a_{0})$ must be out of thermal equilibrium for freeze-in to work, it follows that higher-order derivative interactions must be sufficiently suppressed in order for a clockwork sector to explain freeze-in dark matter. The minimum condition for $\hat{a}_{0}$ to be out of equilibrium is that $\Gamma \lesssim H$ at the reheating temperature $T_{R}$, where $\Gamma$ is the scattering rate of $\hat{a}_{0}$ from thermal bath particles. This also assumes that $T_{R}$ is the highest temperature of the thermal bath, which is true if reheating is instantaneous. For interaction \eqref{eqn:36a}, we can dimensionally estimate the scattering rate to be $\Gamma \sim (O_{10}O_{1j}^{3})^{2}T^{9}/\Lambda_{UV}^{8}$, where we are assuming $\Lambda_{UV} > T$. Therefore, with $H \sim T^{2}/M_{P}$, and including a factor $k_{n}$ to take into account the contribution of other scattering processes when computing the total thermalization rate, $\Gamma_{{\rm total}} = k_{n} \Gamma$ (where we expect that $k_{n} \lesssim 100$), the condition to evade thermalization becomes
\begin{align}
    T_{R} \lesssim  
    \frac{\Lambda_{UV}^{8/7}}{k_{n}^{1/7}(O_{10}O_{1j}^{3})^{2/7}M_{P}^{1/7}}
    \approx
    1\,{\rm TeV}
    \times    \left(
    \frac{100}{k_{n}}
    \right)^{1/7}
    \left(
    \frac{0.3}{(O_{10}O_{1j}^{3})^{2/7}}
    \right)\left(
    \frac{\Lambda_{UV}}{50 \, {\rm TeV}}
    \right)^{8/7} \,.
  \end{align}
For $N$ and $q$ large compared to 1 we find that $O_{10}\approx 1/q$ and $O_{1j} \approx \sqrt{2/N}$ for $j = N/2$. For $N=10$ and $q=4$ these give $(O_{10}O_{15}^{3})^{2/7} \approx 0.3$. The reheating temperature must be larger than the $\mathcal{O}({\rm TeV})$ scale of the clockwork sector, since the heavy clockwork scalars are assumed to be close to relativistic during freeze-in. Therefore it follows that $\Lambda_{UV} \gtrsim 50$ TeV is necessary to allow a window with $T_{R} \gtrsim 1$ TeV. The lower end of this range, which is possible if $T_{R} \sim 1$ TeV and if reheating is instantaneous, is consistent with $\Lambda_{UV}$ being similar to  $f \sim \Lambda \, \sim 10$ TeV in the symmetry-breaking model. However, if $T_{R} \gg 1$ TeV, or if reheating is not instantaneous, then higher-order derivative interactions must be highly suppressed in order for a clockwork sector to explain freeze-in dark matter. This is a strong constraint on the UV origin of the clockwork sector in freeze-in models and suggests that the TeV-scale clockwork sector is effectively renormalizable.

\section{Conclusions}
\label{sec:conclusion} 

We have introduced a TeV-scale clockwork Higgs portal model for scalar dark matter from freeze-in. We have found that freeze-in occurs via the decay of the heavy scalars of the clockwork sector to the Higgs boson and the dark matter scalar, which is the lightest scalar of the clockwork sector. The necessary small portal coupling and dark matter scalar mass can be generated for reasonable values of the global clockwork charge, $q$, and the number of heavy scalars of the clockwork sector, $N$. An interesting feature of the model is that the mass of the dark matter scalar and the strength of the portal coupling are independent of $q$ and $N$ for a given set of model parameters. For a typical TeV-scale clockwork sector, we find that the dark matter scalar has a mass of around 1 MeV. 
 
The clockwork model allows us to understand the very small Higgs portal coupling required for freeze-in purely in terms of the structure of the theory, without the need for any large mass scales. In general, there is no simple symmetry which can eliminate or suppress a coupling between the Higgs boson and a scalar $\phi$ of the form $|H|^2 \phi^2$, and one would dimensionally expect the coupling to be on the order of 1. Therefore, if we wish to avoid simply introducing a very small coupling, a structural explanation such as the clockwork model may be necessary. The model has the added advantage of explaining why the dark matter scalar mass is much less than the scale of electroweak symmetry breaking. 

We note that the clockwork Higgs portal model we have introduced may also be used to generate metastable scalar clockwork weakly interacting massive particle (WIMP) dark matter, along the lines of the fermionic clockwork WIMP dark matter model of \cite{CWDM}.
This is possible if the dark matter WIMP is $\hat{a}_{1}$, which has an unsuppressed Higgs portal interaction and so a conventional WIMP-like freeze-out density, and if its decay to $\hat{a}_{0} + h$ is made extremely slow, by choosing sufficiently large values for $q$ and/or $N$.

For freeze-in to be possible, higher-order derivative interactions between the clockwork scalars must in most cases be highly suppressed in order to prevent thermalization of the dark matter scalars. (An exception to this is the case where the reheating temperature is very low, $T_{R} \sim 1$ TeV, and reheating is instantaneous.)  This is a strong constraint on the UV origin of the clockwork sector. It suggests that the clockwork sector in freeze-in models should be renormalizable below a high UV completion scale. For example, the clockwork sector and the renormalizable Standard Model sector could both originate from a single UV completion at a high energy scale. A renormalizable clockwork sector would also serve as a minimal implementation of the clockwork mechanism.

Finally, we comment on the possibility of experimentally testing this class of model at colliders. The phenomenology of the heavy clockwork scalars will have features in common with the phenomenology of gauge singlet scalars with a $Z_2$ symmetry, which also couple to the SM via the Higgs portal. These are difficult to detect when they can be produced only via off-shell Higgs boson decay. It may be possible to detect their existence at the high-luminosity 14 TeV LHC via their one-loop contribution to the process $pp \rightarrow h^* \rightarrow ZZ$ if the scalar mass is less than around 200 GeV \cite{han}; we anticipate that a similar contribution could arise from heavy clockwork scalars if their mass were not too large compared to the Higgs mass. At future $e^{+} e^{-}$ colliders producing large numbers of Higgs bosons, for example CLIC at 3 TeV, it may be possible to directly produce heavy clockwork scalar pairs via off-shell Higgs decay. The heavy clockwork scalars have a distinctive decay process, where they decay to a lighter clockwork scalar plus either a Higgs boson or, if the mass splitting between the heavy clockwork scalars is less than the Higgs boson mass, a quark or lepton pair via Higgs exchange, with subsequent decay of the lighter clockwork scalars via the same process, ending with missing energy in the form of long-lived next-to-lightest clockwork scalars. This may allow their production to be detected, even if it is at a low rate.

\section*{Acknowledgements} The work of J.M. was partially supported by STFC via the Lancaster-Manchester-Sheffield Consortium for Fundamental Physics.


\end{document}